\documentclass[%
reprint,
superscriptaddress,
amsmath,amssymb,
aps,
pra,
longbibliography,
floatfix,
]{revtex4-2}

\usepackage{graphicx}
\usepackage{dcolumn}
\usepackage{bm}
\usepackage{xcolor}

\newcommand{\ket}[1]{|\rule{0cm}{2ex} #1 \rangle}
\newcommand{\bra}[1]{\langle #1 \rule{0cm}{2ex}|}
\newcommand{\braket}[2]{\bra{#1}#2\rangle}
\newcommand{\avg}[1]{\langle #1 \rangle}
\renewcommand{\Re}{\operatorname{Re}}
\renewcommand{\Im}{\operatorname{Im}}
\newcommand{\sub}[1]{_{\mathrm{#1}}}
\newcommand{\mcA}{\mathcal{A}}
\newcommand{\tx}{\bar{x}}
\newcommand{\tv}{\bar{v}}
\newcommand{\talpha}{\bar{\alpha}}
\newcommand{\tphi}{\bar{\phi}}
\newcommand{\ho}{\ell}
\newcommand{\tvi}{\tilde{v}}
\newcommand{\txi}{\tilde{x}}

\begin{document}
	
\preprint{APS/123-QED}

\title{Beyond the semiclassical approximation in atom interferometry}

\author{W. LaRow}
\affiliation{%
	Department of Physics, University of Virginia,
	Charlottesville, Virginia 22904, USA
}
\author{M. Edwards}
\affiliation{%
        Department of Biochemistry, Chemistry and Physics, 
        Georgia Southern University, Statesboro, Georgia 30460, USA
}
\author{C. A. Sackett}
\email{sackett@virginia.edu}
\affiliation{%
	Department of Physics, University of Virginia,
	Charlottesville, Virginia 22904, USA
}

\date{\today}

\begin{abstract}
We describe a quantum perturbative approach to evaluating the phase shift of an atom interferometer in a weakly anharmonic trap. This provides a simple way to evaluate quantum corrections to the standard semi-classical approximation. The calculation benefits from the use of generalized coherent states for a basis. We find that the form of the semi-classical approximation remains valid to first order in the anharmonic perturbation, but that phase differences arise because the trajectory of a quantum wave packet will generally deviate from that of a classical particle. The quantum correction to the phase is a factor $\ho^2/A^2$ smaller than the semi-classical perturbation itself, where $\ho$ is the quantum harmonic oscillator length scale and $A$ is the classical amplitude of the motion. 
We provide analytical results for one-dimensional perturbations of power three through six in the
position coordinate.
\end{abstract}

\maketitle

\section{Introduction}

Atom interferometry is a useful technique for precision measurements of a variety of phenomena \cite{Berman1997,Cronin2009}. 
Generically,
the sensitivity of an atom interferometer improves as its measurement time and spatial arm separations are
increased. Although most atom interferometers use freely falling atoms, the large fall distance required
for long interrogation times is an experimental challenge \cite{Zhou2011,Kovachy2015,Canuel2018,Zhan2019,Abe2021}. 
One solution is to use 
atoms confined in a trap, and recent experiments have demonstrated good performance with a variety of 
atom-trapping techniques \cite{Weidner2018,Xu2019a,Moan2020,Krzyzanowska2023,Balland2024}.

We focus here on an approach using atoms confined in a harmonic trap, where they
undergo linear or circular oscillations to make up the interferometer trajectory \cite{Moan2020,Krzyzanowska2023,Beydler2024}. 
It is crucial 
in such experiments to understand the effect of the trapping potential itself on the interferometer phase.
In principle this can be achieved through numerical solution of the Schr\"odinger or non-linear Schr\"odinger
equation, as appropriate. However, numerical solution becomes impractical for sufficiently
long measurement times and large arm separations. 
Advances in the technology of atom interferometry for
precision measurements are pushing the field in precisely these directions,
putting direct numerical solution of the Schr\"odinger and non-linear Schr\"odinger equations ever further out of reach.

An alternative tool that is useful
in such situations is the semi-classical approximation (SCA)
\cite{Storey1994,Antoine2003,Antoine2003a,Bongs2006,Hogan2009,Tino2014,Luo2021,Overstreet2021}. 
This assumes that the spatial extent of an atomic wave packet
is negligibly small compared to the length scale over which the external potential varies, but large compared to
the de Broglie wavelength of the propagating matter wave. The approximation is numerically tractable in most cases,
and is simple enough to provide analytical results in some situations.
The semiclassical method is provably exact when the potential energy consists of terms of quadratic order or lower
in the coordinates \cite{Storey1994,Antoine2003,Antoine2003a}. 
A harmonic trap satisfies this condition, but real traps will always include some anharmonicity.
The main goal of this paper is to provide a first-order correction to the SCA
for a trap with weak anharmonic contributions, so that the accuracy of the approximation can be easily assessed.

We find here that the semi-classical phase calculation does in fact remain accurate to first order in 
an anharmonic perturbation, as long as the semi-classical trajectory for a wave packet $x(t)$ 
is replaced by the quantum expectation value $\avg{x}(t)$. 
This expectation value will in general differ at first order from the semi-classical
result, which does therefore cause a quantum correction to the 
phase. For non-interacting atoms, the leading correction is smaller than the SCA phase
by a factor of order $\ho^2/A^2$, where $\ho$ is the quantum harmonic oscillator length scale and
$A$ is the classical amplitude of the motion.

This problem was also recently addressed by Glick and Kovachy \cite{Glick2024}, who developed a field-theoretic method
based on Feynman diagrams. In comparison, our method uses elementary perturbation theory along with
a generalized coherent state basis. The diagrammatic method is more general, applying equally to 
trapped and free-space interferometers and allowing for a larger class of perturbations. 
In contrast, our method more readily provides analytic low-order solutions
for the trapped-atom configuration with power-law anharmonicity.

From an experimental perspective, the SCA has so far been satisfactory, and we are not aware of macro-scale
atom interferometry
measurements that demonstrate a need for quantum corrections. However, as measurement precision improves,
we expect that quantum effects will play a significant role. 
We also suggest that there is a fundamental benefit
in probing effects beyond the SCA as a test of quantum effects on a macroscopic scale. While the 
existence of matter-wave interference is in itself clearly non-classical, alternatives or generalizations
of standard quantum mechanics \cite{Bassi2013,Hu2020} might agree in the semi-classical limit but differ in their
predictions for quantum corrections.

The remainder of the paper is organized as follows: Section \ref{SCAtheory} develops a first-order approximation
for the semi-classical phase, to which the quantum results will be compared. Section \ref{QPT} uses quantum perturbation theory 
to analyze the interferometers. Section \ref{interp} discusses the interpretation of the quantum results and some
ways in which they can be extended. Finally, Section \ref{conc} provides a summary and points to some directions
for future work.

\section{Semi-classical Perturbation Theory} \label{SCAtheory}

\subsection{The semi-classical approximation}

We start with a summary of the conventional SCA for a light-pulse atom interferometer. 
In this approach, the measured phase difference $\theta$ is expressed as 
a sum of three terms \cite{Hogan2009}:
\begin{equation} \label{SCA}
\theta_{SC} = \Delta\phi\sub{prop} + \Delta\phi\sub{laser} + \Delta\phi\sub{sep}.
\end{equation}
The
first is the dynamical propagation phase difference developed by the packets during their time evolution, where
the phase for a single wave packet is given by
\begin{equation} \label{SCAprop}
\phi\sub{prop} = \frac{1}{\hbar} \int_0^t \left[L(t')-E\sub{int}\right]\,dt'.
\end{equation}
Here $L$ is the classical Lagrangian $T-U$ evaluated along the classical trajectory $x(t)$,
and $E\sub{int}$ is the internal energy of the atom based on its quantum state.
This expression conceptually follows 
from the path-integral formulation of quantum mechanics in which the phase associated with a
possible trajectory is the classical action divided by $\hbar$ \cite{Storey1994}; we can interpret the internal
energy $E\sub{int}$ as contributing to the classical potential energy $U$. 

The laser term $\phi\sub{laser}$ comes from the interaction between an atomic wave packet and the laser pulses
used to manipulate it. Every time a laser pulse imparts momentum $p = \hbar \kappa$ to a packet, the
packet also acquires a phase 
\begin{equation} \label{SCAlaser}
\phi\sub{laser} = \kappa x + \xi,
\end{equation}
where $x$ is the position
of the packet at the time of the pulse and $\xi$ is the phase of the laser field relative to an arbitrary fixed
reference. 

The separation phase occurs when the classical trajectories making up the interferometer do not perfectly intersect 
when the packets are recombined. If we consider each packet locally as a plane wave, then the phase gradient of the waves leads
to a phase shift that depends on the packet separation,
\begin{equation} \label{SCAsep}
\Delta\phi\sub{sep} = -\frac{m}{2\hbar}(v_{af}+v_{bf})(x_{bf}-x_{af}),
\end{equation}
where $x_{jf}$ and $v_{jf}$ are the classical position and velocity for packet $j = a,b$ after the final interferometer
light pulse. A non-zero final packet separation will generally lead to a reduction
in the interferometer visibility as well, but the semi-classical approximation does not typically attempt to account for this.

\subsection{Perturbative expansion} 

We apply the SCA method to the problem of an anharmonic oscillator. The atoms move in a one-dimensional potential
\begin{equation}
U(x) = \frac{1}{2} m\omega^2 x^2 + V(x)
\end{equation}
for small perturbation $V(x)$, and we take the internal energy to be zero.
The atoms start at position $x_i$ with velocity $v_i$, to which a laser pulse is applied
to produce a superposition of states with momenta $m v_i + \hbar\kappa_{ai}$ and 
$mv_i + \hbar\kappa_{bi}$. The atoms evolve in the potential for time $t$,
and are then recombined with momentum kicks $\hbar\kappa_{af}$ and $\hbar\kappa_{bf}$.
We seek to evaluate the total SCA phase
\begin{equation}
\theta_{SC} = \phi_{\text{prop,}\,b}-\phi_{\text{prop,}\,a} + \phi_{\text{laser,}\,b}-\phi_{\text{laser,}\,a} + \Delta\phi\sub{sep}.
\end{equation}

For the propagation phase terms, we evaluate
the trajectories to first order in $V$ by taking the position for arm $j=a,b$ to be $x_j(t) = x_{j0}(t) + x_{j1}(t)$,
where $x_{j0}$ is zeroth order in the perturbation and $x_{j1}$ is first order. The various terms 
satisfy the classical equations of motion
\begin{subequations} \label{SCeqm}
\begin{equation} \label{SCeqma}
\ddot{x} + \omega^2 x = -\frac{1}{m} \frac{dV}{dx}
\end{equation}
\begin{equation} \label{SCeqmb}
\ddot{x}_0 + \omega^2 x_0 = 0
\end{equation}
\begin{equation} \label{SCeqmc}
\ddot{x}_1 + \omega^2 x_1 = -\frac{1}{m} \left.\frac{dV}{dx}\right|_{x_0}.
\end{equation}
\end{subequations}
The initial conditions are included in $x_0$, so $x_1(0) = \dot{x}_1(0) = 0$.

It will be convenient to express the interferometer phase in terms of the trajectory variables $x_0$ and $x_1$. 
To this end, consider $\phi\sub{prop}$ for one wave packet:
\begin{equation}
\phi\sub{prop} = \frac{m}{2\hbar} \int_0^t \big(\dot{x}^2 - \omega^2 x^2\big)\,dt' - \frac{1}{\hbar}\int_0^t V(x)\,dt'.
\end{equation}
In the first integral, integrate the $\dot{x}^2$ term by parts to obtain
\begin{equation}
\int_0^t \dot{x}^2\,dt' = x\dot{x} \big|_0^t - \int_0^t \ddot{x} x\,dt'.
\end{equation}
Then using \eqref{SCeqma} to substitute for $\ddot{x}$ and, approximating $V$ to first order, 
we have
\begin{align}
\phi\sub{prop} & = \frac{m}{2\hbar}[x(t)\dot{x}(t)-x(0)\dot{x}(0)]  \nonumber\\
   & \quad + \frac{1}{\hbar}\int_0^t \left[\frac{x_0}{2}  \left.\frac{dV}{dx}\right|_{x_0}- V(x_0)\right]\,dt'.
\end{align}
Substitute from \eqref{SCeqmc} to relate
\begin{equation} \label{integral1}
\int_0^t x_0 \left.\frac{dV}{dx}\right|_{x_0}\,dt' = -m\int_0^t x_0 \left(\ddot{x}_1 + \omega^2 x_1\right)\,dt'.
\end{equation}
Integrating by parts twice yields
\begin{equation}
\int_0^t x_0 \ddot{x}_1 \,dt' = \left.x_0 \dot{x}_1\right|_0^t - \left.x_1 \dot{x}_0\right|_0^t + \int_0^t \ddot{x}_0 x_1\,dt',
\end{equation}
and then application of \eqref{SCeqmb} results in cancellation of the $\int \omega^2 x_0 x_1\,dt$ term in \eqref{integral1}.
Using also the initial conditions for $x_1$, we finally
obtain
\begin{align} \label{SCprop}
\phi\sub{prop}  & = \frac{m}{2\hbar}\big[x(t)v(t)-x(0)v(0) - x_0(t)v_1(t) + x_1(t) v_0(t)\big] \nonumber\\
& \quad - \frac{1}{\hbar}\int_0^t V(x_0)\,dt'
\end{align}
where we relabel $\dot{x} \rightarrow v$.

To apply this result to the interferometer phase $\theta$, we set $x_a(0) = x_i$ and $v_a(0) = v_i + \hbar\kappa_{ai}/m$,
and we abbreviate the final coordinates as $x_a(t) = x_a$, $v_a(t) = v_a$. This gives
\begin{align}
\phi_{\text{prop,}a} & = \frac{m}{2\hbar}(x_a v_a - x_i v_i - x_{a0}v_{a1} + x_{a1} v_{a0}) -\frac{1}{2}\kappa_{ai}x_i \nonumber\\
& \quad - \frac{1}{\hbar}\int_0^t V[x_{a0}(t')]\,dt',
\end{align}
with a similar expression for $\phi_{\text{prop,}b}$. For the laser phase, we have
\begin{equation}
\phi_{\text{laser,}b}-\phi_{\text{laser,}a} = \kappa_{bi}x_i - \kappa_{ai}x_i + \kappa_{bf}x_b -  \kappa_{af}x_a + \xi,
\end{equation}
with $\xi \equiv \xi_b -\xi_a$ representing a phase shift of the recombination pulse.
The separation phase is
\begin{align}
\Delta\phi\sub{sep} & 
    = -\frac{m}{2\hbar}\left(v_a + \frac{\hbar\kappa_{af}}{m} +v_b + \frac{\hbar\kappa_{bf}}{m}\right)\big(x_b-x_a\big) \nonumber \\
    & = \frac{m}{2\hbar}\big(x_av_a - x_bv_b + x_av_b-x_bv_a \big) \nonumber \\
     & \qquad   + \frac{1}{2}\big(\kappa_{af}x_a - \kappa_{bf}x_b + \kappa_{bf}x_a- \kappa_{af}x_b \big).
\end{align}
Summing all these terms, we obtain
\begin{align} \label{SCtheta}
\theta_{SC} & = \xi + \frac{m}{2\hbar}(x_a v_b - x_b v_a) 
    + \frac{1}{2}(\kappa_{bi} - \kappa_{ai})x_i \nonumber \\
    & \quad + \frac{1}{2}(\kappa_{bf}-\kappa_{af})(x_a+x_b) \nonumber \\
    & \quad + \frac{m}{2\hbar}\big[x_{b0}v_{b1} - x_{b1} v_{b0}-x_{a0}v_{a1} + x_{a1} v_{a0}\big] \nonumber \\
    & \quad -\frac{1}{\hbar}\int_0^t [V(x_{b0})-V(x_{a0})]\,dt'.
\end{align}
We can then separate out the perturbative orders, writing $\theta_{SC} = \theta_0 + \theta_1$
for
\begin{align}
\theta_0 & = \xi + \frac{m}{2\hbar}(x_{a0} v_{b0} - x_{b0} v_{a0})  + 
    \frac{1}{2}(\kappa_{bi} - \kappa_{ai})x_i \nonumber \\
    & \quad + \frac{1}{2}(\kappa_{bf}-\kappa_{af})(x_{a0}+x_{b0})
\end{align}
and
\begin{align} \label{theta1SC}
\theta_1 & = \frac{m}{2\hbar}\big[(v_{b0}-v_{a0})(x_{a1}+x_{b1})-(v_{a1}+v_{b1})(x_{b0}-x_{a0})\big] \nonumber \\
   & \quad +\frac{1}{2}(\kappa_{bf}-\kappa_{af})(x_{a1}+x_{b1}) \nonumber \\
   & \quad -\frac{1}{\hbar}\int_0^t [V(x_{b0})-V(x_{a0})]\,dt'.
\end{align}

In an experiment, the interferometer output would typically be detected by measuring the 
population $P_f$ in the final momentum state, which varies as
\begin{equation}
P_f = \frac{1}{2}\big(1+\mathcal{V}\cos\theta\big)
\end{equation}
for visibility $\mathcal{V}$. The atomic contribution to $\theta$ could
be determined by, for instance, scanning $\xi$ to produce an interference fringe.

\subsection{Cubic potential}

To illustrate the general results above, we apply them to a trap with cubic anharmonicity $V(x) = \beta x^3$.
Although practical traps are more likely to exhibit perturbations with even powers of $x$, the asymmetry
of the cubic potential means that its effects can be observed in a straightforward
double-Bragg interferometer geometry,
which simplifies the example here.

We consider an interferometer with atoms starting at rest at the trap center, $x_i = v_i = 0$. The atoms are split symmetrically with
$\kappa_{ai} = -\kappa_{bi} = \kappa$, undergo nominally one half-period of oscillation with $\omega t = \pi$,
and are then recombined to rest with $\kappa_{af} = -\kappa_{bf} = \kappa$. 
Solving the equations of motion \eqref{SCeqm} yields
\begin{subequations} \label{SCx}
\begin{equation}  \label{SCx0}
x_{a0}(t) = A\sin \omega t
\end{equation}
\begin{equation}
x_{a1}(t) = -\frac{\beta A^2}{2m\omega^2}\big(3-4\cos\omega t + \cos2\omega t\big),
\end{equation}
\end{subequations}
where $A = \hbar\kappa/m\omega$ is the unperturbed amplitude of the motion. 
The result for $x_b$ can be obtained by reversing the sign of $A$, so at $\omega t = \pi$ we have
$x_{a0} = x_{b0} = 0$, $v_{a0} = -v_{b0} = -A\omega$,
$x_{a1} = x_{b1} =  -4\beta A^2/m\omega^2$, and $v_{a1} = v_{b1} = 0$. 
We also have
\(
\int_0^{\pi} \sin^3 u\,du = 4/3.
\)
Using these results, we find $\theta_0 = 0$ and the interferometer phase is
\begin{equation} \label{SCcubicphase}
\theta_{SC} = \frac{8}{3} \frac{\beta A^3}{\hbar\omega}.
\end{equation}
This arises entirely from the integral term in Eq.~\eqref{theta1SC}, since the terms involving $(v_{b0}-v_{a0})$ and
$(\kappa_{bf}-\kappa_{af})$ cancel, while the $(v_{a1}+v_{b1})$ term is zero.

\begin{figure}
\includegraphics[width=3in]{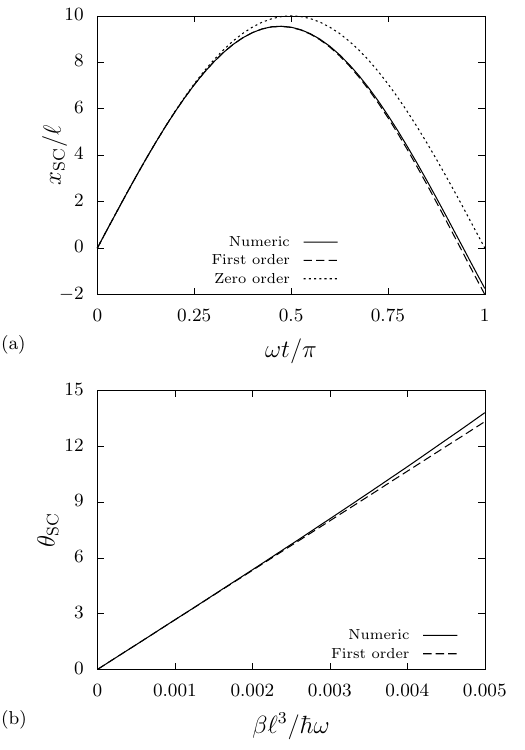}
\caption{\label{SCfig} Results of the semi-classical approximation for a symmetric interferometer in
an anharmonic trap with perturbation $V(x) = \beta x^3$. 
Lengths are scaled by $\ho = \sqrt{\hbar/m\omega}$. (a) Trajectory of a classical particle
starting with a velocity kick $\omega A$ for unperturbed oscillation amplitude $A = 10\ho$, with 
$\beta = 0.005 \hbar\omega/\ho^3$. The solid
line is a numerical solution of the exact equation of motion \protect\eqref{SCeqma}. The dashed
line is the perturbative result of Eq.~\protect\eqref{SCx}. The dotted line shows the trajectory
for the unperturbed oscillator, Eq.~\protect\eqref{SCx0}, for comparison. 
(b) Interferometer phase $\theta_{SC}$ from the semi-classical approximation. The solid line shows
the numerical solution, and the dashed line shows the perturbative result of \protect\eqref{SCcubicphase}.
}
\end{figure}

Figure~\ref{SCfig} compares the results for the trajectory and phase to an exact numerical 
solution of the classical equations of motion. The perturbative result \eqref{SCcubicphase} correctly
captures the leading dependence on $\beta$ of the exact calculation.

\section{Quantum perturbative method} \label{QPT}

\subsection{General approach}

We now seek to develop an analogous perturbative solution starting from a fully quantum description of the motion.
We consider non-interacting atoms confined in a nearly harmonic potential, with Hamiltonian $H = H_0 + V(x)$ where
\begin{equation}
H_0 = \frac{p^2}{2m} + \frac{1}{2} m\omega^2 x^2
\end{equation}
describes a harmonic oscillator and $V(x) \propto x^\lambda$ 
is a weak anharmonic perturbation that we assume to have power-law form with integer $\lambda>2$. 
Here we consider a one-dimensional system, but we discuss in Section \ref{interp} how the technique
might be extended to higher dimensions. We also discuss there how mean-field interactions might be incorporated
in an approximate way.

Our approach uses the standard formalism of time-dependent perturbation theory in the interaction picture
\cite{Shankar1994}. The
time evolution of the Schr\"odinger-picture wave function is given by
\begin{equation}
\psi(t) = U_0(t) U_I(t) \psi(0),
\end{equation}
where $U_0(t)$ is the unperturbed time-evolution operator $\exp(-i H_0 t/\hbar)$ and $U_I(t)$ is the evolution operator
in the interaction picture with respect to $V$. This satisfies the perturbation expansion
\begin{align} \label{expansion}
U_I(t) & =  1 - \frac{i}{\hbar} \int_0^t dt_1 V_I(t_1) \nonumber \\
& \quad - \frac{1}{\hbar^2} \int_0^t dt_2 \int_0^{t_2} dt_1 V_I(t_2) V_I(t_1) + \ldots,
\end{align}
where the perturbation in the interaction picture is
\begin{equation}
V_I(t) = U_0^\dagger(t) V U_0(t).
\end{equation}
Our strategy is to express $\psi(t)$ in a basis of generalized coherent states, and then evaluate
the effect of $U_I$ using matrix mechanics.

\subsection{Generalized coherent states}

The generalized coherent (GC) states $\ket{\alpha,n}$ have seen
occasional use in quantum optics contexts \cite{Senitzky1954,Boiteux1973,Philbin2014}. They are defined by 
\begin{equation}
\ket{\alpha,n} = \hat{D}(\alpha) \ket{n},
\end{equation}
where $\ket{n}$ is the harmonic-oscillator eigenstate with energy $(n+1/2)\hbar\omega$ and
$\hat{D}(\alpha)$ is the unitary displacement operator \cite{Mandel1995}
\begin{equation}
\hat{D}(\alpha) = e^{\alpha \hat{a}^\dagger - \alpha^* \hat{a}}.
\end{equation}
Here $\hat{a}$ and $\hat{a}^\dagger$ are the usual harmonic oscillator ladder operators, and 
$\alpha$ is an arbitrary complex number.
Conceptually, the GC state is a copy of the energy eigenstate $\ket{n}$ that
oscillates in the harmonic potential with a complex amplitude proportional to $\alpha$. 
For fixed $\alpha$, the set $\left\{\ket{\alpha,n}\right\}$ forms a
complete orthonormal basis~\cite{Philbin2014} that we will use for the perturbation expansion.

The $\ket{\alpha,0}$ state is identical
to the ordinary coherent state $\ket{\alpha}$ satisfying $\hat{a}\ket{\alpha} = \alpha\ket{\alpha}$.
To understand the action of $\hat{a}$ on the generalized state $\ket{\alpha,n}$, use the 
commutator $[\hat{a},\hat{D}(\alpha)] = \alpha D$ and thus
\begin{align}
\hat{a} \ket{\alpha,n} & =  \hat{a} \hat{D}(\alpha)\ket{n} = \big[\alpha \hat{D}(\alpha) + \hat{D}(\alpha) \hat{a}\big]\ket{n} \nonumber \\
& = \alpha \ket{\alpha,n} + \sqrt{n}\ket{\alpha,n-1}.
\end{align}
In a similar way,
\begin{equation}
\hat{a}^\dagger \ket{\alpha,n} = \alpha^* \ket{\alpha,n} + \sqrt{n+1}\ket{\alpha,n+1}.
\end{equation}
With these, we obtain the action of the position operator,
\begin{align} \label{xoperation}
\hat{x}\ket{\alpha,n} & \equiv  \frac{\ho}{\sqrt{2}}\left(\hat{a}+\hat{a}^\dagger\right)\ket{\alpha, n} \nonumber \\
& = x_c \ket{\alpha, n} + \frac{\ho}{\sqrt{2}}\!\left[ \sqrt{n} \ket{\alpha,n\!-\!1} + \sqrt{n\!+\!1} \ket{\alpha,n+1}\right],
\end{align}
where $\ho \equiv \sqrt{\hbar/m\omega}$ and $x_c \equiv \sqrt{2}\ho \Re\alpha$. Noting that $\braket{\alpha,m}{\alpha,n} = \delta_{mn}$,
we see that the expectation value of the position is $\langle x \rangle = x_c$. We similarly find
the action of the momentum operator
\begin{align} \label{p_operator}
\hat{p}\ket{\alpha,n} & = mv_c \ket{\alpha,n} \nonumber \\
& \quad +i \frac{m\omega\ho}{\sqrt{2}}\left[ \sqrt{n+1} \ket{\alpha,n+1} - \sqrt{n} \ket{\alpha,n-1}\right].
\end{align}
where $v_c \equiv \sqrt{2}\omega \ho \Im \alpha$ is the expectation value of the velocity.

The time evolution of the GC states follows from that of the eigenstates $\ket{n}$ and the
operators $\hat{a}$, $\hat{a}^\dagger$. From $U_0(t) \ket{n} = e^{-i(n+1/2)\omega t}\ket{n}$
and $U_0(t) D(\alpha) U_0(-t) = D(\alpha e^{-i\omega t})$, we obtain
\begin{equation}
U_0(t) \ket{\alpha, n} = e^{-i(n+1/2)\omega t} \ket{\alpha e^{-i\omega t},n}.
\end{equation}
This time dependence for $\alpha$ means that the center position follows the classical trajectory
$x_c(t) = x_c(0) \cos\omega t + [v_c(0)/\omega] \sin\omega t$ and that $v_c(t) = dx_c/dt$.

The spatial wave function $\psi_{\alpha,n}(x) \equiv \braket{x}{\alpha,n}$ can be determined using
the harmonic oscillator eigenstate wave functions and the translation properties of the displacement operator, 
yielding \cite{Senitzky1954}
\begin{align} \label{psix}
\psi_{\alpha,n}(x,t) & = \frac{\mcA e^{i\phi_c}}{\sqrt{2^n n!}}  H_n\!\left(\sqrt{\frac{m\omega}{\hbar}}(x-x_c)\right) \nonumber\\
& \quad \times \exp\left[-\frac{m\omega(x-x_c)^2}{2\hbar} + i\frac{mv_c}{\hbar}(x-x_c)\right].
\end{align}
Here $\mcA = (m\omega/\pi\hbar)^{1/4}$, $H_n$ is the $n$th Hermite polynomial, and as above,
$x_c = \sqrt{2}\ho\Re\alpha$ and $v_c = \sqrt{2}\ho\omega\Im \alpha$.
The central phase $\phi_c = \arg[\psi(x_c)]$ is given by
\begin{align} \label{phic}
\phi_c(t) & = \frac{m}{2\hbar} \left\{ x_c(0) v_c(0) + \int_0^t \left[v_c(t')^2 - \omega^2 x_c(t')^2\right]\,dt' \right\} \nonumber\\
& \quad -\left(n+\frac{1}{2}\right) \omega t .
\end{align}
The time evolution of $\phi_c$ is consistent with the semi-classical propagation phase in Eq.~\eqref{SCAprop}, taking $E\sub{int} = (n+1/2)\hbar\omega.$ This correspondence is expected
since the SCA is exact for a harmonic potential. As in the semi-classical calculation, the integral in \eqref{phic} can be evaluated by 
parts to obtain the alternative form $\phi_c = m x_c(t) v_c(t)/2\hbar -(n+1/2)\omega t$.

From Eq.~\eqref{p_operator} with $\hat{p}\rightarrow -i\hbar (\partial/\partial x)$, we can also obtain
the derivative
\begin{equation} \label{GCderiv}
\frac{\partial \psi_{\alpha,n}}{\partial x} = i\frac{mv_c}{\hbar} \, \psi_{\alpha,n} + \frac{1}{\sqrt{2}\ho} \left[ \sqrt{n}\, \psi_{\alpha,n-1} 
-\sqrt{n+1}\, \psi_{\alpha,n+1}\right],
\end{equation}
which will be used below. A final relation we require is
\begin{equation} \label{overlap}
\braket{\alpha,n}{\beta,0} = \braket{\alpha}{\beta} \frac{(\beta-\alpha)^n}{\sqrt{n!}},
\end{equation}
where $\braket{\alpha}{\beta} = \exp[-|\alpha-\beta|^2/2]\exp[(\alpha^*\beta-\beta^*\alpha)/2]$ is the 
overlap integral for standard coherent states. Result~\eqref{overlap} is readily derived from Eq.~(1.6) of 
Ref.~\cite{Boiteux1973}.

\subsection{First-order solution}

Returning to the perturbation calculation, we take the initial state $\ket{\psi(0)}$ of the atoms to be the GC state 
$\ket{\alpha_i,0}$, corresponding to a non-interacting Bose condensate that has potentially been displaced by
a laser pulse.
We express the subsequent evolution as a perturbation series
\begin{equation} \label{psiexp}
\ket{\psi(t)} = \ket{\psi_0(t)} + \ket{\psi_1(t)} + \ldots
\end{equation}
where $\ket{\psi_\mu}$ scales with $V^\mu$. We have then
\begin{equation}
\ket{\psi_0(t)} = U_0(t) \ket{\psi(0)} = e^{-i\omega t/2} \ket{\alpha(t),0}
\end{equation}
for $\alpha(t) = \alpha_i e^{-i\omega t}$. We express the first-order correction as 
\begin{equation} \label{psi1}
\ket{\psi_1(t)} = \sum_{n}  D_{n}(t) e^{-i\omega t/2}\ket{\alpha(t), n}
\end{equation}
with coefficients
\begin{equation} \label{Dmn}
D_{n} = -\frac{i}{\hbar} e^{-in\omega t} \int_0^t dt_1 e^{i\omega n t_1} \bra{\alpha(t_1),n}V\ket{\alpha(t_1),0}
\end{equation}
following from Eq.~\eqref{expansion}.

The matrix elements in \eqref{Dmn} are evaluated in closed form in the Appendix,
and are polynomials in the unperturbed position $x_c(t_1)$. 
The integrand in \eqref{Dmn} thus involves products of
sines and cosines in addition to the explicit complex exponential. These can always be integrated analytically,
with examples provided in the Appendix.

This method gives an expression for the wave function at time $t$. To relate to the semi-classical
result, it is useful to evaluate the expectation values $\tx \equiv \avg{x}$ and $\tv \equiv \avg{v}$
for the position and velocity of the particle. 
To first order in $V$ we have
\begin{equation} \label{xm}
\tx(t) = \bra{\psi}\hat{x}\ket{\psi} = x_c(t) + \sqrt{2}\ho \Re D_{1}(t),
\end{equation}
and similarly
\begin{equation} \label{vm}
\tv(t) = \frac{1}{m} \bra{\psi}\hat{p}\ket{\psi} = v_c(t) + \sqrt{2}\ho\omega \Im D_{1}(t).
\end{equation}
We express these in perturbation orders as $\tx = x_0 + x_1$ and $\tv = v_0 + v_1$, identifying
$x_0 = x_c = \sqrt{2}\ho \Re \alpha(t)$ and $v_0 = v_c = \sqrt{2}\ho \Im \alpha(t)$. Evidently, we can 
introduce a perturbed coherent state parameter 
\begin{equation} \label{tilde}
\talpha = \alpha + D_{1} = \frac{1}{\sqrt{2}\ho}\left(\tx + i\frac{\tv}{\omega}\right)
\end{equation} such that
$\ket{\talpha}$ gives the correct position and velocity for the perturbed evolution.

\subsection{Interferometer analysis}

Using the perturbation results, we consider the operation of an interferometer analogous 
to that of Section~\ref{SCAtheory}. The initial coherent state is given by $\alpha_i = (x_i + i v_i/\omega)/\sqrt{2}\ho$. 
A laser pulse generates momentum kicks $\hbar \kappa_{ai}$
and $\hbar \kappa_{bi}$, to produce a superposition. 
The laser also imparts a phase shift; to model this, we take the effect of the
laser as multiplication by $e^{i\kappa x}$, which from \eqref{psix} and \eqref{phic}
generally results in
\begin{equation}
e^{i\kappa x}\ket{\alpha,n} = e^{i\kappa x_c/2}\left|\alpha+i\kappa\ho/\sqrt{2},n\right\rangle.
\end{equation}
The state after the beam-splitting operation is therefore
\begin{equation}
\ket{\psi_i} = \frac{1}{\sqrt{2}} \left[ e^{i\kappa_{ai} x_i/2} \ket{\alpha_{ai},0}Z
    + e^{i\kappa_{bi} x_i/2} \ket{\alpha_{bi},0} \right],
\end{equation}
for $\alpha_{ji} = [x_i + i (v_i/\omega +\kappa_{ji}\ell^2)]/\sqrt{2}\ell$ with $j = a,b$.

Each of the wave packets then evolves for time $t$ such that, per Eq.~\eqref{psi1},
\begin{equation}
\ket{\alpha_{ji},0} \rightarrow \ket{\alpha_j,0} + \sum_n D_{jn} \ket{\alpha_j,n}.
\end{equation}
Here we write $D_{jn}$ for the $D_{n}$ coefficient corresponding to packet $j$,
and $\alpha_j = \alpha_{ji} e^{-i\omega t} \equiv (x_j+iv_j/\omega)/\sqrt{2}\ho$.

The laser is now applied again as a recombination pulse. We consider the final
state $\ket{\psi_f}$ produced by applying laser kick $\kappa_{af}$ to packet $a$ with phase $\xi_a$, and
kick $\kappa_{bf}$ to packet $b$ with phase $\xi_b$. The resulting state is
\begin{align}
\ket{\psi_f} & = \frac{1}{2} \Big[ 
    e^{i[(\kappa_{ai}x_i + \kappa_{af}x_a)/2 + \xi_a]} \left(\ket{\alpha_{af},0} + \sum D_{an} \ket{\alpha_{af},n}\right) \nonumber \\
& \quad + e^{i[(\kappa_{bi}x_i + \kappa_{bf}x_b)/2 + \xi_b]} \left(\ket{\alpha_{bf},0} + \sum D_{bn} \ket{\alpha_{bf},n}\right) \Big] \nonumber \\
   & \equiv \frac{1}{2}\left( \ket{\psi_{af}}  + \ket{\psi_{bf}}\right),
\end{align}
with $\alpha_{jf} = \alpha_j + i\kappa_{jf}\ho/\sqrt{2}$. 

In some cases it is necessary to track the population in all final states, for instance if
multiple interference paths can occur \cite{Weitz1996,Lu2018}. For the simple interferometer considered here,
however, we need only evaluate the population in state $f$, given by 
\begin{equation} \label{interfere}
\braket{\psi_f}{\psi_f} = \frac{1}{4}\left[2 
    +  \braket{\psi_{af}}{\psi_{bf}} + \braket{\psi_{bf}}{\psi_{af}}\right].
\end{equation}
Using Eq.~\eqref{overlap} and expanding to first order in the perturbation, we have
\begin{widetext}
\begin{equation}
\braket{\psi_{af}}{\psi_{bf}}  = e^{i[(\kappa_{bi}x_i + \kappa_{bf}x_b - \kappa_{ai}x_i - \kappa_{af}x_a)/2 + \xi]}\braket{\alpha_{af}}{\alpha_{bf}} 
 \left\{1 + 
    \sum_n \frac{1}{\sqrt{n!}}\left[D_{an}^* (\alpha_{bf}-\alpha_{af})^n + D_{bn} (\alpha_{af}^*-\alpha_{bf}^*)^n\right]\right\},
\end{equation}
\end{widetext}
with $\xi = \xi_b-\xi_a$.
To obtain interference, $\alpha_{af}$ and $\alpha_{bf}$ should be similar. We suppose $|\alpha_{af} -\alpha_{bf}| \ll 1$
and truncate the result to first order in this quantity. We can then also approximate
\begin{equation}
\braket{\alpha_{af}}{\alpha_{bf}} \approx 1 + \frac{1}{2}\left(\alpha_{af}^*\alpha_{bf} - \alpha_{af}\alpha_{bf}^*\right)
= 1 + i\Im[\alpha_{af}^*\alpha_{bf}].
\end{equation}
The final population in \eqref{interfere} becomes $(1 + \cos\theta)/2$
with phase
\begin{align}
\theta = & ~ \xi +  \frac{1}{2}\big(\kappa_{bi}x_i + \kappa_{bf}x_b - \kappa_{ai}x_i - \kappa_{af}x_a\big)  \nonumber \\
    & + \Im(\alpha_{af}^*\alpha_{bf}) + \Im\left[D_{a0}^* + D_{b0} \right] \\
    & +\Im \left[D_{a1}^*\alpha_{bf}-D_{a1}^*\alpha_{af} + D_{b1}\alpha_{af}^*-D_{b1}\alpha_{bf}^*\right].\nonumber 
\end{align}
The interferometer visibility here is equal to 1 because of the first-order approximation for $|\alpha_{bf}-\alpha_{af}|$.

To simplify the phase expression further, note that $\kappa x/2 = \Im(i\ho\kappa\alpha/\sqrt{2} )$. 
If we also apply $\alpha_{jf} = \alpha_j + i\kappa_{jf}\ho/\sqrt{2}$, we can express $\theta$ compactly as
\begin{align} \label{theta}
\theta = \Im\Big[ & \talpha_{af}^* \talpha_{bf} +
    i\ho\big(\kappa_{bi}\alpha_i - \kappa_{ai}\alpha_i + \kappa_{bf}\talpha_{b} -\kappa_{af}\talpha_a\big)/\sqrt{2} \nonumber \\
    &  + D_{b0}-D_{a0} +D_{b1}^*\alpha_b - D_{a1}^*\alpha_a\Big] + \xi,
\end{align}
where we use the perturbed parameters of \eqref{tilde} with $\talpha_j = \alpha_j + D_{j1}$ and 
$\talpha_{jf} = \talpha_j + i\kappa_{jf}\ho/\sqrt{2}$. 
We can also use Eq.~\eqref{Dmn} to express
\begin{align}
\Im D_{j0} & = -\frac{1}{\hbar} \int_0^t  \bra{\alpha_j(t_1),0}V\ket{\alpha_j(t_1),0}\, dt' \nonumber\\
& \equiv -\frac{1}{\hbar} \int_0^t \avg{V}_j\, dt'
\end{align}
as the integral of the expectation value of the perturbation over the unperturbed trajectory.
In this way, we obtain the final result in
terms of position and velocity parameters as
\begin{align} \label{qmtheta}
\theta = & ~\xi +  \frac{m}{2\hbar}(\tx_a \tv_b - \tx_b \tv_a) 
    + \frac{1}{2}(\kappa_{bi} - \kappa_{ai})x_i \nonumber \\
   & + \frac{1}{2}(\kappa_{bf}-\kappa_{af})(\tx_a+\tx_b) \nonumber \\
   &  +\frac{m}{2\hbar}(x_{b1}v_{b0} - v_{b1}x_{b0} - x_{a1}v_{a0}+v_{a1}x_{a0}) \nonumber \\
   & -\frac{1}{\hbar} \int_0^t\big[\avg{V}_b-\avg{V}_a\big]\,dt'.
\end{align}
Here we have set $\talpha_j = (\tx_j + i \tv_j/\omega)/\sqrt{2}\ho$, $\alpha_j = (x_{j0} + i v_{j0}/\omega)/\sqrt{2}\ho$, and
$D_{j1} = (x_{j1} + i v_{j1}/\omega)/\sqrt{2}\ho$, as respectively the total, zeroth-order, and first-order components 
of the quantum trajectory.

Notably, \eqref{qmtheta} has the same form as Eq.~\eqref{SCtheta} from the SCA, if we identify
the quantum expectation values $\tx$, $\tv$, and $\avg{V}$ with the classical variables $x, v$,
and $V(x)$. In this sense, we conclude that the quantum result is consistent (to first order) 
with the semi-classical prescription \eqref{SCA}.
However, this does not mean that the
semi-classical phase result is correct, because in general the quantum trajectory $\tx(t)$ will
differ from the classical $x(t)$, and $\avg{V}(t)$ differs from the classical $V(x(t))$. 
Ehrenfest's theorem \cite{Shankar1994} ensures that
$m\,d\avg{v}/dt = -\avg{dU/dx}$
but if the potential $U$ contains powers higher than quadratic, we do not generally have  
$\avg{dU/dx} = dU/d\avg{x}\big|_{\avg{x}}$. This difference is illustrated in the example below.

\subsection{Evaluation for cubic anharmonicity} \label{qcubic}

The method described above is applicable to any power law perturbation. 
We illustrate it here for a cubic anharmonicity $V(x) = \beta x^3$, as the simplest non-trivial example.
We consider atoms starting with $x_i = 0$ and $v_i = A\omega$, for nominal
oscillation amplitude $A$. Taking $x_c(t) = A\sin\omega t$, we use Eqs.~\eqref{appD30} and \eqref{appD31}
from the Appendix with $\txi = 0$ and $\tvi = A/\ell$:
\begin{align}
D_{0} & = i\frac{\beta A}{\hbar\omega} 
    \bigg[\frac{A^2}{12}\left(-8+9\cos\omega t - \cos 3\omega t\right) \nonumber \\
    & \qquad \qquad - \frac{3}{2}\ho^2(1-\cos\omega t)\bigg]  \\
D_{1} & = \frac{\beta \ho}{\sqrt{2}\hbar\omega} 
    \bigg[ \frac{A^2}{4}\left(-3 e^{-2i\omega t}+8e^{-i\omega t} -6 + e^{2i\omega t}\right) \nonumber \\
       & \qquad \qquad  +\frac{3}{2}\ho^2\left(e^{-i\omega t}-1\right)\bigg]  
\end{align}
The mean position is then
\begin{align} \label{cubicxm}
\tx & = x_c + \sqrt{2}\ho\Re D_{1} \nonumber \\
& = A\sin\omega t -\frac{1}{2} \frac{\beta A^2}{m\omega^2}(3-4\cos\omega t+\cos 2\omega t) \nonumber \\
& \quad - \frac{3}{2} \frac{\hbar \beta}{m^2\omega^3} (1-\cos\omega t).
\end{align}
Comparing to the semi-classical result of Eq.~\eqref{SCx}, we see that there is an additional term representing
the quantum correction to the motion.

Applying this result to a symmetric interferometer with $\kappa_{ai} = \kappa_{af} = -\kappa_{bi} = -\kappa_{bf} = m\omega A/\hbar$
and $\omega t = \pi$, we find from \eqref{qmtheta} that
\begin{equation} \label{qmthetacubic}
\theta = \frac{8}{3} \frac{\beta A^3}{\hbar\omega} + 6 \frac{\beta A}{m\omega^2}.
\end{equation}
Again the leading term for large $A$ agrees with the semi-classical prediction of Eq.~\eqref{SCcubicphase},
but there is a quantum correction term that is smaller by a factor of order $\ho^2/A^2$.

\begin{figure}
\includegraphics[width=3.5in]{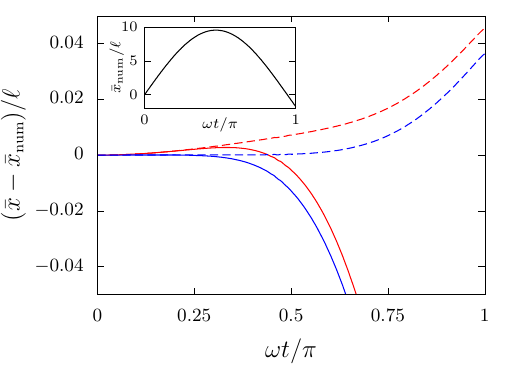}
\caption{\label{qtraj} (Color online)
Trajectories $\tx(t)$ in an anharmonic trap with perturbation $V = \beta x^3$, for $\beta = 0.005\hbar\omega/\ho^3$
and initial velocity $10\omega \ho$ with $\ho = \sqrt{\hbar/m\omega}$. The inset shows the expectation value
$\tx$ calculated from exact numerical solution of the Schr\"odinger equation. The main plot shows
the deviation from the exact result for the various approximations considered: the first-order
semi-classical result (solid red), the second-order semi-classical result (red dashes),
the first-order quantum perturbation result (solid blue), and the second-order quantum result (blue dashes).
}
\end{figure}

We compare these results to both the semi-classical approximation and exact numerical solution of the Schr\"odinger equation.
For the exact solution, we solve the Schr\"odinger equation using an explicit staggered-time
algorithm \cite{Visscher1991}. Figure~\ref{qtraj} shows the discrepancy between the $\avg{x}$ values
from the numerical solution and the predictions of the first-order quantum perturbation calculations (solid blue)
and the SCA (solid red).
It is evident that the quantum perturbation result remains accurate for longer. 
To better check the theory, we carried out the perturbation calculations to second
order (blue and red dashes).
For this, evaluation of the perturbation matrix elements in Eq.~\eqref{expansion}
involves six applications of the $\hat{x}$ operator at two different times, making the calculation
less tractable. However, with the aid of Mathematica for symbolic manipulations, 
we obtained analytical results for the second-order contributions to $\avg{x}$, 
\begin{align}
x_2 = & \frac{1}{16} \frac{\beta^2 A^3}{m^2 \omega^4} (-60 \omega t \cos \omega t + 5 \sin \omega t \nonumber \\
& \qquad \qquad + 32 \sin 2\omega t - 3 \sin 3\omega t)  \nonumber \\
& + \frac{5}{2} \frac{\beta^2 \hbar A}{m^3\omega^5}(-3 \omega t \cos \omega t + \sin\omega t + \sin 2\omega t).
\end{align}
We emphasize the first-order results in our development as they are simpler and
we expect the quantum corrections to be small in most practical
situations, but the good agreement at second order does validate the GC method.

\begin{figure}
\includegraphics[width=3.5in]{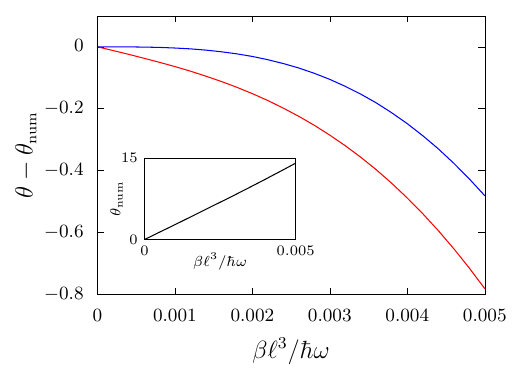}
\caption{\label{qphase} (Color online)
Interferometer phase $\theta$ for a symmetric interferometer in an anharmonic trap with 
perturbation $V = \beta x^3$, for packets with initial velocity $\pm 10\omega \ho$. 
The laser phase $\xi$ is zero. The black curve is calculated using exact numerical solutions
for the Schr\"odinger equation. The blue curve is the first-order quantum perturbation result
of Eq.~\eqref{qmthetacubic}. The red curve is the first-order semi-classical result of \protect\eqref{SCcubicphase}.
}
\end{figure}

Figure~\ref{qphase} shows the exact and first-order approximations for the interferometer phase 
$\theta$, as a function of $\beta$. The black curve is the exact result computed as
$\arg(\braket{\psi_a}{\psi_b})$ where $\psi_a$ and $\psi_b$ are the final numerical wave functions
for the two packets of the interferometer. The blue line shows the perturbative result of
Eq.~\protect\eqref{qmthetacubic}, and the red line shows the semi-classical result of \protect\eqref{SCcubicphase}.
It is evident that the quantum perturbation result improves upon the semi-classical approximation.

\section{Interpretation and Discussion} \label{interp}

The parameters used in Figs.~\ref{qtraj} and \ref{qphase} were chosen to illustrate the quantum
corrections in a regime where the exact Schr\"odinger equation is 
possible to solve numerically, enabling us to check the accuracy of the perturbation results.
The calculations for Fig.~\ref{qphase} nonetheless took two hours to run using a fast algorithm
on a new desktop PC. The run time scales as $(A/\ell)^4$, which quickly becomes intractable as
$A$ increases. For example, in future experiments we hope to achieve 
$A = 300 \ho$, corresponding to about 2 mm for Rb atoms in a 3-Hz trap. 
Numerical solution for these conditions would require substantial computational resources,
and while this is perhaps achievable
in one dimension, extension to three dimensions seems beyond reach.
In contrast, the analytical method presented here 
can readily be extended to three-dimensions using a product GC basis
$\ket{\alpha_x,n_x}\ket{\alpha_y,n_y}\ket{\alpha_z,n_z}$.  
As long as the analysis is restricted to first order, 
the effects of the perturbation on each degree of freedom can
be analyzed independently and the results presented here can be directly applied. 

For a realistic interferometer with
multiple laser pulses, the wave function evolution must be evaluated in stages between pulses. The results derived here
assume that the wave function at the start of a stage is a GC state with $n=0$, which is not generally
the case after the first stage. However, it is accurate at first order to replace the perturbed wave function
$\ket{\psi} = \ket{\psi_0} + \ket{\psi_1}$ at the end of a stage with the displaced state
$\ket{\bar{\alpha},0}$, using $\bar{\alpha}$ from Eq.~\eqref{tilde}. In this way, arbitrary interferometer
sequences can be evaluated by the method presented here. Alternatively, the assumption
that the initial quantum state has the form $\ket{\alpha,0}$ can be relaxed. In that case,
it would be necessary to consider more general matrix elements $\langle \alpha, n | V |\alpha, m\rangle$.

The results here are also useful for qualitative estimation. We see 
that the semi-classical component of the phase perturbation will generically be of order
$\avg{V}/\hbar\omega$ where $\avg{V}$ is the average of the perturbation over the classical trajectory. The quantum correction
to this phase correction will be smaller by a factor of order $\ho^2/A^2$, 
for classical amplitude $A$. For a perturbation $V \propto x^\lambda$, the quantum correction will
scale as $A^{\lambda-2}$, so for $\lambda > 2$, 
the SCA becomes relatively more accurate as $A$ becomes large, but the absolute size of the quantum correction increases without bound. As an example, the experiments of Refs.~\cite{Luo2021,Beydler2024} used $^{87}$Rb atoms moving through
a complicated trajectory in
a trap with $\omega = 2\pi\times 3.5$~Hz and motional amplitude $A \approx 80\ho$. One relevant
anharmonicity was quartic, $V = \beta_4 x^4$, with a coefficient $\beta_4 \approx (-2\times 10^{-7}) \hbar\omega/\ho^4$.
Without the need for a detailed calculation, we can 
estimate the semi-classical phase perturbation to be of order $\beta_4 A^4/\hbar\omega \sim 10$~rad with a quantum
correction of order $\beta_4 A^2 \ho^2/\hbar\omega \approx 10^{-3}$~rad. 
This is small enough to neglect, but 
if the amplitude is increased to $300 \ell$ as proposed above, the corrections would become
significant for a precision measurement. Interactions can also be expected to
increase the quantum corrections, as discussed below.

The consistency between the quantum phase expression of Eq.~\eqref{qmtheta} and the semi-classical result \eqref{SCtheta}
suggests that the semi-classical idea is applicable even in the anharmonic case. This seems reasonable, since the path-integral
formulation of quantum mechanics is itself valid for potentials of any form. It is worth noting, however, that we cannot in general
identify the propagation phase with the phase of the wave function at even the quantum-corrected position $\tx$.
From
\eqref{qmtheta}, we can subtract $\Delta\phi\sub{laser}$ and $\Delta\phi\sub{sep}$ to get an expression
for the perturbed propagation phase on a given path,
\begin{equation} \label{phipropQ}
\phi_{\text{prop}} =  \frac{m}{2\hbar} (x_0 v_0 -x_i v_i + 2x_1 v_0) + \Im D_{0}.
\end{equation}
However, if we evaluate 
$\tphi \equiv \arg[\psi(\tx)]$, we obtain a different result. Using $\tx = x_0 + x_1$
from Eq.~\eqref{xm}, we expand
\begin{equation}
\psi(x_0+x_1) \approx \psi_0(x_0) + x_1 \left.\frac{\partial\psi_0}{\partial x}\right|_{x_0} + \psi_1(x_0).
\end{equation}
Since $x_0 = x_c$, the first and last terms can be evaluated using Eq.~\eqref{psix} as
\begin{equation}
\psi_0(x_0) = \mcA e^{i\phi_c} \qquad 
\psi_1(x_0) = \mcA e^{i\phi_c} \sum_{n} h_n D_{n} \label{psi1x0},
\end{equation}
where $\phi_c = (m/2\hbar)(x_{0}v_0-x_i v_i)$ is the central phase for the unperturbed system and
the coefficients
\begin{equation}
h_n = \frac{H_n(0)}{\sqrt{2^n n!}} = \begin{cases}
\left(-\frac{1}{2}\right)^{n/2} \binom{n}{n/2}^{1/2} & (n~\text{even}) \\
0 & (n~\text{odd})
\end{cases}
\end{equation}
come from $\psi_{\alpha,n}(x_c)$ in Eq.~\eqref{psix}.
The derivative term is evaluated with the aid of Eq.~\eqref{GCderiv} as
\begin{equation}
\left.\frac{\partial\psi_0}{\partial x}\right|_{x_0} = i \frac{mv_0}{\hbar} \mcA  e^{i\phi_c}.
\end{equation}
These combine to give
\begin{equation}
\psi(\tx) \approx \mcA e^{i\phi_c} \left[ 1 + i\frac{m}{\hbar}v_0 x_1
    + \sum_{\text{even}~n} h_n D_{n} \right].
\end{equation}
Interpreting the term in brackets as an expansion of $e^{i\phi_1}$, we obtain the central phase
\begin{equation} \label{perturbphi}
\tphi = \frac{m}{2\hbar} (x_0 v_0 - x_i v_i + 2x_1v_0) + \sum_{\text{even}~n} h_n \Im{D_{n}}.
\end{equation}
This is similar to Eq.~\eqref{phipropQ}, but the sum includes terms with $n>0$ that do not contribute to
the interferometer phase. The validity of these terms was confirmed by comparison to numerical
solution of the Schr\"odinger equation. We conclude that although the SCA concept remains useful
for anharmonic systems, it is not possible to interpret the quantities involved as directly as
in the harmonic case.

Another feature of our result is that the difference between the SCA and quantum phases
is determined by the quantum corrections to the trajectory $\tx(t)$ and expectation value $\avg{V}$. 
These corrections
arise because the quantum wave function samples the perturbation over a range of values, and not 
just at a single classical point. In the case of the cubic perturbation, the quantum
result can be recovered by replacing the classical potential $\beta x^3$ with an effective
potential averaged over a Gaussian wave function:
\begin{align}
V\sub{eff}(x) & = \frac{1}{\ho\sqrt{4\pi}}\int_{-\infty}^{\infty} e^{-(x-x')^2/\ho^2} \beta x'^3\,dx' \nonumber\\
& = \beta \left(x^3 + \frac{3}{2}\ho^2 x\right),
\end{align}
where the Gaussian corresponds to the probability distribution over $x'$ of an ordinary coherent state centered at $x$.
If we use this potential in the SCA calculation,
we obtain agreement with the quantum results. However, for perturbations with power four or higher,
this simple replacement does not suffice.

A more systematic way to approach this problem is with the Lagrangian variational method (LVM) \cite{PerezGarcia1996,Thomas2022}.
Here we suppose a trial wave function that depends on various parameters, and we find 
equations of motion for those parameters by constructing an effective classical Lagrangian
and applying the Euler-Lagrange technique. The calculation using $V\sub{eff}$ above is equivalent
to the LVM for a Gaussian trial wave function with $x'$ as a variational parameter. In the case of
a quartic perturbation, the quantum results can be recovered using a Gaussian where both the
center and width are variational parameters. We plan to further explore this application of the LVM to atom
interferometry in future work.

The effective potential or LVM approaches could also be useful for incorporating the effects of interactions
into the phase calculation. An interacting condensate will typically have a Thomas-Fermi wave function with
size $L \gg \ho$ \cite{Dalfovo1999}. Mean-field interactions can be included directly in the LVM Lagrangian, with the size $L$
adopted as a variational parameter. Alternatively, an effective potential like $V\sub{eff}$ can be defined using a Thomas-Fermi,
rather than a Gaussian, probability distribution. For the purpose of estimation, it should normally be 
adequate to simply substitute $\ho \rightarrow L$ in the quantum perturbation results. For instance,
the interferometer of \cite{Beydler2024} described above used condensates of about $2\times 10^4$ atoms,
which gives a Thomas-Fermi size $L \approx 3\ho$. The quantum correction to the anharmonic phase
should therefore be estimated as $\beta_4 A^2 L^2/\hbar\omega \approx 10^{-2}$~rad, about ten times
larger than expected for the non-interacting case and already potentially significant.

\section{Summary and Future Work} \label{conc}

We have developed a quantum perturbation method appropriate for atom interferometers in 
nearly harmonic traps. Generalized coherent states provide an efficient basis for the calculation 
and lead to expressions for the required matrix elements that are easy to compute. We have shown
that, at least to the first order of accuracy, the semi-classical approximation for phase estimation
can be used in an anharmonic trap, but that the wave-packet trajectories are generally 
modified by quantum effects. The quantum correction is generally smaller than the 
semi-classical result by a factor of $\ho^2/A^2$, where $\ho$ is the quantum harmonic
oscillator length scale and $A$ is the classical amplitude of motion.

This work raises a number of questions for further investigation. A notable one is whether
and when the general semi-classical approach breaks down entirely, in the sense that propagation,
laser, and separation phases cannot be usefully defined and evaluated. As a start, it would be 
interesting to extend the general result of Eq.~\eqref{qmtheta} to second order and compare
to the corresponding semi-classical expression. It may also be necessary to consider
more carefully the light-atom interaction process: it is generally not  
accurate to treat the light field as an ideal plane wave \cite{LouchetChauvet2011,Schkolnik2015,Zhou2016}, 
and as the size of a wave
packet increases, the atoms will sample the optical deviations more strongly.

Another useful extension would be a method to evaluate the leading-order quantum
correction to the phase while incorporating the exact classical motion.
It is easy to envision practical situations where the anharmonicity is
large enough that a perturbative expansion of the classical trajectory
is inadequate, but at the same time the quantum corrections are small enough
to approximate in first order. The effective potential and LVM approaches might
be ways to achieve this goal, but it might also be possible to achieve a partial
resummation of the quantum perturbation expansion \eqref{expansion} to express the
classical part in all orders while retaining quantum corrections to the
desired degree of accuracy.

It may also be possible to extend the analysis to situations more general
than trapped atom interferometers with power-law perturbations. Free-space
solutions can be obtained from our results in the limit
$\omega\rightarrow 0$, but in that case the initial wave packet of an interferometer
is not usually close to the ground state of the potential.
Also, more general perturbations such as a localized `bump' would 
typically lead to the coupling of many generalized coherent states, making
our method inefficient. These situations might be more easily handled by
diagrammatic methods \cite{Glick2024}.

In a practical direction, we hope to apply the techniques developed
here to the three-dimensional interferometer configuration of Ref.~\cite{Beydler2024},
and in particular to evaluate the impact on the semi-classical results
of Ref.~\cite{Luo2021}. The scaling analysis described above indicates that the
quantum corrections will be fairly small, but large enough to be of 
potential significance in high-performance applications.

Finally, it would be interesting to measure a quantum correction effect
in an interferometry experiment. For a trapped-atom experiment, our results suggest that the optimum
configuration would use weak harmonic confinement (providing large $\ho$) and a 
small amplitude of motion (low $A$), along with a relatively strong anharmonic
perturbation to produce a measurable phase. We hope that the work presented here will stimulate efforts 
in this area.

\vspace{2ex}

\begin{acknowledgments}
The authors gratefully acknowledge helpful conversations with Jonah Glick, Simon Haine, and Maxim Olshanii. We also thank Jeff Heward for help with the simulations. C.A.S. thanks the Quantum Science and Technology group at the Australian National University for hosting
him while portions of this work were carried out.
This work was supported by the National Science Foundation, Grant Nos.~2110471 (C.A.S. and W.L.)
and 2207476 (M.E.).

\end{acknowledgments}

\vspace{0.3in}

\appendix*

\section{}

We derive here the coefficients $D_n$ of Eq.~\eqref{Dmn}, which can be expressed as
\begin{equation}
D_{n} = -\frac{i\beta_\lambda}{\hbar} e^{-in\omega t} \int_0^t e^{i\omega n t_1} f_n(t_1) \,dt_1 
\end{equation}
for matrix elements
\begin{equation}
f_n(t_1) = \bra{\alpha(t_1),n}x^\lambda\ket{\alpha(t_1),0}
\end{equation}
with perturbation $V = \beta_\lambda x^\lambda$ and integer $\lambda$. We will first show that
\begin{equation} \label{derivf}
f_n(t_1) = \frac{\eta}{\sqrt{n}} \frac{d}{dx_c} f_{n-1}(t_1),
\end{equation}
where $x_c = \eta[\alpha(t_1) + \alpha^*(t_1)]$ and $\eta = \ell/\sqrt{2} = \sqrt{\hbar/2m\omega}$. 
We then derive a general expression for $f_0$, from 
which $f_n$ can be obtained by differentiation. 

To obtain \eqref{derivf} we start with
\begin{equation}
\langle\alpha, n|\hat{x}^{\lambda}|\alpha,0\rangle 
=
\langle 0,n|\hat{D}^{\dag}(\alpha)\hat{x}^{\lambda}\hat{D}(\alpha)|0,0\rangle,
\end{equation}
and we apply the displacement operator relation
\begin{equation} \label{displacement}
\hat{D}^{\dag}(\alpha)\hat{x}^{\lambda}\hat{D}(\alpha) =
\left(\hat{x} + x_{c}\right)^{\lambda}.
\end{equation}
This gives
\begin{align}
f_n & = \langle 0,n|\left(\hat{x} + x_c\right)^{\lambda}|0,0\rangle \nonumber \\
& = \frac{1}{\sqrt{n}}
\langle n-1|\hat{a}\left(\hat{x} + x_c\right)^{\lambda}|0\rangle
\label{first_eq}
\end{align}
where $\ket{0,n} \equiv \ket{n}$ is the ordinary oscillator eigenstate and we have used
$\bra{n-1}\hat{a} = \sqrt{n}\bra{n}$. Since $x_c$ is a scalar and 
$\hat{x} = \eta(\hat{a}^\dagger +\hat{a})$, we can expand
\begin{equation}
\left(\hat{x} + x_{c}\right)^{\lambda}
=
\sum_{k=0}^{\lambda}\binom{\lambda}{k}\eta^{k}x_{c}^{\lambda-k}
\left(\hat{a}^{\dag} + \hat{a}\right)^{k}
\end{equation}
and obtain
\begin{equation} \label{thirdeq}
f_n
=
\frac{1}{\sqrt{n}}
\sum_{k=0}^{\lambda}\binom{\lambda}{k}\eta^{k}x_{c}^{\lambda-k}
\langle n-1|\hat{a}\left(\hat{a}^{\dag} + \hat{a}\right)^{k}|0\rangle.
\end{equation}
We now apply the commutator relation
\begin{equation} 
\left[
\hat{a},\left(\hat{a}+\hat{a}^{\dag}\right)^{k}
\right] = 
k\left(\hat{a}+\hat{a}^{\dag}\right)^{k-1}
\label{comm1}
\end{equation}
to see that
\begin{equation}
\langle n-1|\hat{a}  \left(\hat{a}^{\dag} + \hat{a}\right)^{k}|0\rangle 
= k\langle n-1|\left(\hat{a}+\hat{a}^{\dag}\right)^{k-1}|0\rangle 
\end{equation}
since $\hat{a}|0\rangle = 0$. This then gives
\begin{equation}
f_n = \frac{1}{\sqrt{n}}
\sum_{k=1}^{\lambda}\binom{\lambda}{k}k\eta^{k}x_{c}^{\lambda-k}
\langle n-1|\left(\hat{a}^{\dag} + \hat{a}\right)^{k-1}|0\rangle
\label{fourtheq}
\end{equation}
where we can start the sum at $k = 1$ since the $k=0$ term is zero.
Finally, we use the identities
\begin{equation}
\binom{\lambda}{k}k = \binom{\lambda}{k-1}(\lambda-k+1)
\end{equation}
and
\begin{equation}
(\lambda-k+1)\eta^{k}x_{c}^{\lambda-k} = 
\eta\frac{d}{dx_{c}}
\left(
\eta^{k-1}x_{c}^{\lambda-k+1}
\right),
\end{equation}
to obtain
\begin{align}
f_n & = \frac{\eta}{\sqrt{n}} \frac{d}{dx_c}
\sum_{k=1}^{\lambda}\binom{\lambda}{k\!-\!1}\eta^{k-1}x_{c}^{\lambda-k+1}
\langle n\!-\!1|\left(\hat{a}^{\dag}\!\! +\! \hat{a}\right)^{k-1}\!|0\rangle \nonumber \\
& = \frac{\eta}{\sqrt{n}} \frac{d}{dx_c}
\sum_{k'=0}^{\lambda}\binom{\lambda}{k'}\eta^{k'}x_{c}^{\lambda-k'}
\langle n-1|\left(\hat{a}^{\dag} + \hat{a}\right)^{k'}|0\rangle \nonumber \\
& = \frac{\eta}{\sqrt{n}} \frac{d}{dx_c} f_{n-1}
\end{align}
as desired.
In the second equality above, we have shifted the summation index by setting $k^{\prime}=k-1$ and extended the top summation limit from $k^{\prime}=\lambda-1$ to $k^{\prime}=\lambda$. This does not change the value of the right-hand side because the $k^{\prime}=\lambda$ term is independent of $x_{c}$ and thus its derivative with respect to $x_{c}$ is zero.

We now consider $f_0 = \langle \alpha, 0 | \hat{x}^\lambda |\alpha, 0\rangle \equiv p_\lambda$.
As above, we have
\begin{align}\label{pn1}
p_\lambda & = \langle 0 | (\hat{x}+x_c)^\lambda|0\rangle \nonumber \\
& = \sum_{k=0}^\lambda \binom{\lambda}{k} \eta^k x_c^{\lambda-k} \langle 0 | (\hat{a} + \hat{a}^\dagger)^k|0\rangle.
\end{align}
Let us define
\begin{equation}
q_k = \langle 0 | (\hat{a} + \hat{a}^\dagger)^k|0\rangle.
\end{equation}
By parity, $q_k = 0$ for odd $k$, and it is easy to show that $q_0 = q_2 = 1$. We will derive
a recursion relation for $q_k$.

We have
\begin{align} \label{qeqn1}
q_{k+2} & = \langle 0 | (\hat{a}+\hat{a}^\dagger)^2 (\hat{a} + \hat{a}^\dagger)^k|0\rangle \nonumber \\
    & = \langle 0 |(\hat{a}\hat{a} + \hat{a}\hat{a}^\dagger + \hat{a}^\dagger\hat{a} + \hat{a}^\dagger\hat{a}^\dagger)(\hat{a} + \hat{a}^\dagger)^k|0\rangle 
\end{align}
Since $\langle 0|\hat{a}^\dagger = 0$ and $\langle 0|\hat{a}\hat{a}^\dagger = \langle 0|$, \eqref{qeqn1}
reduces to 
\begin{equation}
q_{k+2} = \langle 0 | \hat{a}^2 (\hat{a} + \hat{a}^\dagger)^k|0\rangle + q_k.
\end{equation}

For the remaining term, use the commutator \eqref{comm1} twice, to get
\begin{align}
q_{k+2} & = k(k-1)\langle 0 |(\hat{a} + \hat{a}^\dagger)^{k-2}|0\rangle + q_k \nonumber \\
    & = k(k-1)q_{k-2} + q_k.
\end{align}
This recursion, along with the $q_0$ and $q_2$ values, gives
\[
q_k = (k-1)(k-3)...(1) = (k-1)!!.
\]
Since $k$ is even, we can use the identity
\begin{equation}
(k-1)!! = \frac{k!}{2^{k/2} (k/2)!}.
\end{equation}
Applying this to \eqref{pn1} and writing the binomial coefficient as factorials, we have
\begin{equation}
p_\lambda = 
\sum_{\text{even } k=0}^\lambda \frac{\lambda!}{2^{k/2} (k/2)!(\lambda-k)!} \eta^k x_c^{\lambda-k}, \label{pn2}
\end{equation}
giving $p_\lambda$ as a polynomial in $x_c$. Some results for relevant $\lambda$ are
\begin{align}
p_3(x_c) & = x_c^3 + 3\eta^2 x_c \\
p_4(x_c) & = x_c^4 + 6 \eta^2 x_c^2 + 3 \eta^4 \\
p_5(x_c) & = x_c^5 + 10 \eta^2 x_c^3 + 15 \eta^4 x_c \\
p_6(x_c) & = x_c^6 + 15 \eta^2 x_c^4 + 45 \eta^4 x_c^2 + 15 \eta^6
\end{align}

The coefficients $D_n(t)$ are obtained by first computing $f_n(t_1)$, and then integrating 
$e^{-i\omega t_1} f_n(t_1)$. In general, $x_c(t_1) = x_c(0) \cos(\omega t_1) + [v_c(0)/\omega] \sin(\omega t_1)$,
so the integrals involve various products of $\sin\omega t_1$ and $\cos\omega t_1$. These can always
be evaluated analytically using standard techniques.
General results for $\lambda = 3$ to 6 are listed below,
with the notation $D_{\lambda n} \equiv D_n$ for perturbation power $\lambda$ and we use
dimensionless variables $\txi = x_c(0)/\ho$, $\tvi = v_c(0)/\omega\ho$: 
\begin{widetext}
\begin{align} \label{appD30}
D_{30}(t) = \frac{i\beta_3\ell^3}{12\hbar\omega}\Big[ &-2\tvi(9+ 4 \tvi^2 +6 \txi^2) 
        + 9\tvi(2  + \tvi^2  +  \txi^2 )\cos\omega t 
     - \tvi(\tvi^2 - 3 \txi^2) \cos 3\omega t \nonumber \\
        & - 9\txi(2   +\tvi^2   + \txi^2) \sin\omega t 
        + \txi(3 \tvi^2  - \txi^2) \sin 3\omega t\Big]
\end{align}
\begin{equation} \label{appD31}
D_{31}(t) = \frac{\beta_3\ell^3}{4\sqrt{2}\hbar\omega} \Big[  -3 (\tvi - i \txi)^2 e^{-2i\omega t}  
     +  2(3 + 4 \tvi^2 - 4 i \tvi \txi + 2 \txi^2 )e^{-i\omega t} 
     - 6 (1 + \tvi^2 + \txi^2)
     +  (\tvi + i \txi)^2 e^{2i\omega t} \Big]
\end{equation}
\begin{align}
D_{40}(t) = -\frac{i\beta_4\ell^4}{32\hbar\omega}\big[ & 4\tvi \txi( 12  + 3 \tvi^2  + 5 \txi^2)
     + 12(2 + 4 \tvi^2 + 4 \txi^2 + \tvi^4 + 2 \tvi^2 \txi^2 + \txi^4)\omega t  
   -16\tvi \txi( 3 + \tvi^2 + \txi^2 ) \cos 2\omega t\nonumber \\
  & +4\tvi \txi( \tvi^2 - \txi^2)\cos 4\omega t
   -8 (\tvi^2-\txi^2)(3  + \tvi^2  + \txi^2  ) \sin 2\omega t
   + (\tvi^4  - 6 \tvi^2 \txi^2  + \txi^4) \sin 4\omega t  \big]
\end{align}
\begin{align}
D_{41}(t) = -\frac{i\beta_4\ell^4}{8\sqrt{2}\hbar\omega}\Big[ &
    2 (\tvi - i \txi)^3  e^{-3i\omega t}
    + ( 12 \tvi + 12 i \txi + 3 \tvi^3 + 9 i \tvi^2 \txi + 15 \tvi \txi^2 + 5 i \txi^3 ) e^{-i\omega t} \nonumber \\             
    & + 12 i (\tvi-i\txi)(2 + \tvi^2 + \txi^2 ) \omega t  e^{-i\omega t} 
     - 6 (\tvi + i \txi)(2 + \tvi^2 + \txi^2))  e^{i\omega t} 
     + (\tvi + i\txi)^3   e^{3i\omega t}
 \Big]
\end{align}
\begin{align}
D_{50}(t) = -\frac{i\beta_5 \ell^5}{240\hbar\omega}\Big[ & 
    4 \tvi(225 + 200 \tvi^2+ 300 \txi^2  + 32 \tvi^4 + 80 \tvi^2 \txi^2 + 60 \txi^4) 
    -150 \tvi (6 +  6 \tvi^2 + 6 \txi^2 + \tvi^4 + 2 \tvi^2 \txi^2 + \txi^4 ) \cos \omega t \nonumber \\
   & + 25 \tvi (4 \tvi^2- 12 \txi^2 +\tvi^4 - 2 \tvi^2 \txi^2   -  3 \txi^4 ) \cos 3\omega t 
    -3 \tvi (\tvi^4 - 10 \tvi^2 \txi^2 + 5 \txi^4 ) \cos 5\omega t \nonumber \\
   & + 150 \txi (6 + 6 \tvi^2 + 6 \txi^2 + \tvi^4 + 2 \tvi^2 \txi^2 + \txi^4)\sin\omega t 
    - 25 \txi ( 12 \tvi^2- 4 \txi^2 + 3 \tvi^4  +2 \tvi^2 \txi^2 - \txi^4) \sin 3\omega t \nonumber \\
   & + 3 \txi ( 5 \tvi^4 -10 \tvi^2 \txi^2 + \txi^4) \sin 5\omega t
   \Big]
\end{align}
\begin{align}
D_{51}(t) = \frac{\beta_5\ell^5}{48\sqrt{2}\hbar\omega}\Big[ &
    5 (\tvi - i \txi)^4  e^{-4i\omega t}
    - 60 (\tvi - i \txi)^2 (3 + \tvi^2 + \txi^2)  e^{-2i\omega t} \nonumber \\
    & + 4 (45 + 120 \tvi^2  - 120 i \tvi \txi + 60 \txi^2 + 32 \tvi^4 - 32 i \tvi^3 \txi  + 48 \tvi^2 \txi^2 - 48 i \tvi \txi^3 + 12 \txi^4 ) e^{-i\omega t} \nonumber \\
    & - 90 (2 + 4\tvi^2 + 4\txi^2 + \tvi^4 + 2\tvi^2\txi^2 + \txi^4) \nonumber \\
    & + 20 (\tvi + i \txi)^2 (3 + \tvi^2 + \txi^2) e^{2i\omega t}
    - 3 (\tvi + i \txi)^4 e^{4i\omega t}
    \Big]
\end{align}
\begin{align}
D_{60}(t) = -\frac{i\beta_6 \ell^6}{192\hbar\omega}\Big[ &
    4 \tvi \txi ( 270 + 135 \tvi^2 + 225 \txi^2 + 15 \tvi^4 + 40 \tvi^2 \txi^2 + 33 \txi^4 )  \nonumber \\
    & + 60 (6 + 18 \tvi^2 + 18 \txi^2 + 9 \tvi^4 + 18 \tvi^2 \txi^2 + 9 \txi^4 
            + \tvi^6 + 3 \tvi^4 \txi^2 + 3 \tvi^2 \txi^4 + \txi^6 ) \omega t  \nonumber \\
    & - 90 \tvi \txi (12 + 8 \tvi^2 + 8 \txi^2 + \tvi^4 + 2 \tvi^2 \txi^2 + \txi^4) \cos(2\omega t)
    + 36 \tvi \txi (\tvi^2 - \txi^2)(5  + \tvi^2 - \txi^2) \cos(4\omega t) \nonumber \\
    & - 2 \tvi \txi ( 3 \tvi^4 - 10 \tvi^2 \txi^2 + 3 \txi^4 ) \cos(6\omega t)
    - 45  (\tvi^2-\txi^2) ( 12 + 8 \tvi^2 + 8 \txi^2+ \tvi^4 + 2 \tvi^2 \txi^2 + \txi^4 ) \sin(2\omega t)  \nonumber \\
    & + 9 (5 \tvi^4 - 30 \tvi^2 \txi^2 + 5 \txi^4 
            + \tvi^6 - 5 \tvi^4 \txi^2 - 5 \tvi^2 \txi^4 + \txi^6 ) \sin(4\omega t) \nonumber \\
    & - (\tvi^2-\txi^2)(\tvi^4 - 15 \tvi^2 \txi^2) \sin(6\omega t)
    \Big]
\end{align}
\begin{align}
D_{61}(t) = \frac{i\beta_6 \ell^6}{64\sqrt{2}\hbar\omega} \Big[ &
    3 (\tvi - i \txi)^5  e^{-5i\omega t}
    - 30 (\tvi-i\txi)^3 (4 + \tvi^2 + \txi^2)  e^{-3i\omega t}
    - 4  (90 \tvi + 90 i \txi 
            + 45 \tvi^3 + 135 i \tvi^2 \txi + 225 \tvi \txi^2 + 75 i \txi^3  \nonumber \\
            & \qquad \qquad + 5 \tvi^5 + 25 i \tvi^4 \txi + 40 \tvi^3 \txi^2 + 40 i \tvi^2 \txi^3 + 55 \tvi \txi^4 + 11 i \txi^5 ) e^{-i\omega t}  \nonumber \\
    & - 120 i (\tvi - i \txi)(6 + 6 \tvi^2 + 6 \txi^2  + \tvi^4 + 2 \tvi^2 \txi^2 + \txi^4) \omega t            e^{-i\omega t} \nonumber \\
    & + 60 (\tvi + i \txi)(6 + 6 \tvi^2 + 6 \txi^2 + \tvi^4 + 2 \tvi^2 \txi^2 + \txi^4) e^{i\omega t} \nonumber \\
    & - 15 (\tvi + i \txi)^3 (4 + \tvi^2 + \txi^2) e^{3i\omega t} 
    + 2 (\tvi + i \txi)^5  e^{5i\omega t}
\Big]
\end{align}

\end{widetext}  

%

\end{document}